\documentclass[epjCONF,columns]{svjour} % for 2 columns format
\usepackage{graphics}
\usepackage[varg]{txfonts} % Times fonts
\usepackage[latin1]{inputenc}
\usepackage{lineno}
\usepackage{amssymb}
\session-title{Hadron Collider Physics Symposium 2011}
\begin{document}
%
%\linenumbers
%
\title{Hard QCD Results with Jets at the LHC}
\author{Sven Menke\inst{1}\fnmsep\thanks{\email{menke@mppmu.mpg.de}}
  on behalf of the ATLAS and CMS collaborations}
\institute{Max-Planck-Institut f{\"u}r Physik, F{\"o}hringer Ring 6,
  80805 M{\"u}nchen, Germany}
\abstract{ Hard QCD results in proton-proton collisions at
  $\sqrt{s}=7\,{\rm TeV}$ with jets from data recorded up to the end
  of 2010 by the CMS and ATLAS experiments at the LHC are
  reported. Inclusive jet and di-jet cross section measurements as
  well as observables sensitive to multi-jet activity are shown and
  compared to simulations based on leading log parton showers as well
  as NLO QCD predictions. Novel approaches to identify highly boosted
  massive final states by exploiting the jet substructure are tested
  on the dominant QCD background.  } %end of abstract
\maketitle
\section{Introduction}
\label{Intro}
The ATLAS~\cite{art:ATLAS} and CMS~\cite{art:CMS} experiments both
have rich QCD programs involving high $p_\perp$ jets aiming to probe
the structure of the colliding protons, to measure the strong coupling
constant and to test the standard model (SM) at the shortest distance
scales accessible today in the high-center-of-mass proton-proton
collisions of the LHC. Furthermore deviations from the SM would
indicate the presence of new physics beyond the SM. The datasets
recorded up to the end of 2010 and corresponding to integrated
luminosities of ${\cal L}\simeq35\,{\rm pb}^{-1}$ per experiment have
been used by ATLAS and CMS to update their measurements of inclusive
jet and di-jet cross sections, as well as to measure multi-jet and
angular di-jet distributions.  New methods based on the sub-structure
of jets to detect heavily boosted massive objects ending up in single
jets at the LHC have also been tested on the dominant QCD background
and compared to expectations.

\section{Jet reconstruction and calibration}
\label{JetReco}
The infrared- and collinear-safe Anti-$k_\perp$ jet clustering
algorithm~\cite{art:antiKt} is used by both experiments in the
inclusive reconstruction mode with distance parameters $0.4\le
R\le0.7$. Input to the jet algorithm are 4-vectors stemming either
from stable particles in generator-level simulations, partons in NLO
calculations, topological calorimeter
clusters~\cite{art:TopoCluster,art:ATLASPerf} in ATLAS or particle
flow (PF) objects~\cite{art:ParticleFlow,art:ParticleFlowII} in CMS in
full simulations and data. Topological clusters can be calibrated
prior to the jet making~\cite{art:ATLASPerf} in ATLAS or left at the
electromagnetic (EM) scale. The PF objects use information from all
CMS subsystems and are calibrated to correspond to stable particles
like $\gamma$'s, leptons, charged and neutral hadrons. In all cases
residual jet-level corrections are needed to account for particle
losses not detectable on cluster or PF object
level~\cite{art:CMSJES,art:ATLAS-CONF-2011-032}
with larger corrections (up to a factor of $2$) for EM-scale inputs
and small corrections (on the level of $5-10\%$) for already
calibrated inputs. The jet-level calibrations are Monte Carlo (MC)
based correction functions in $|\eta|$ and $p_\perp$. Jet energy scale
(JES) and uncertainty are validated with in-situ methods using
$p_\perp$ balance in di-jet and $\gamma$-jet events and the momentum
projection fraction method in $\gamma$-jet events. The Monnte Carlo
based correction factors are validated with single particle test-beam
data and $E/p$ measurements of isolated hadrons in collision data
which are then extrapolated using fragmentation predictions to the
jet-level.  The systematic JES uncertainty is typically $3-6\%$ for
both ATLAS and CMS over a large range of pseudo-rapidities and
$p_\perp$, with the larger values at large $|\eta|$, very low and very
high $p_\perp$.

\section{Inclusive jet cross section measurements}
\label{XSec}
The inclusive jet cross section is measured by
ATLAS~\cite{art:ATLAS-CONF-2011-047} and
CMS~\cite{art:PhysRevLett.107.132001} as a function of transverse jet
momentum $p_\perp$ and jet rapidity $y$.
The data is corrected bin-by-bin for migration effects in $p_\perp$
due to the steeply falling spectrum in $p_\perp$ and the finite
\begin{figure}[htb]
\begin{center}
\centerline{%
  \resizebox{0.25\textwidth}{!}{\includegraphics{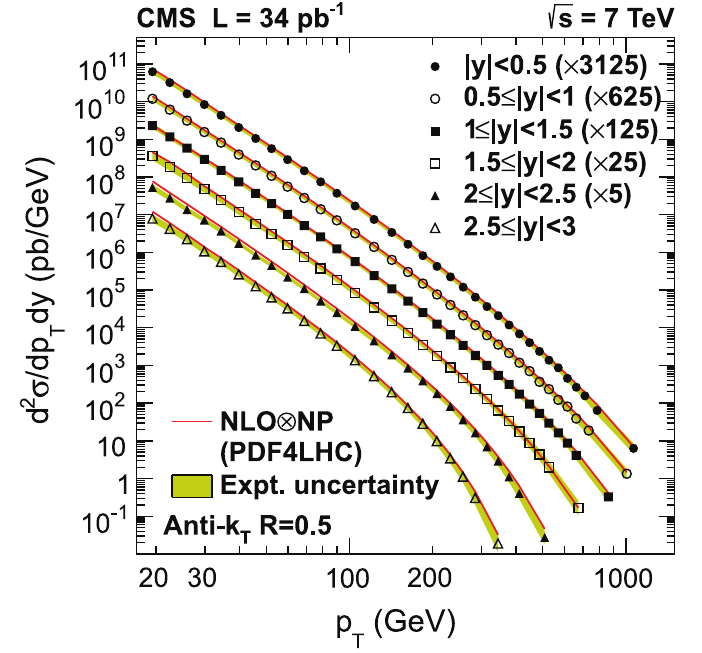}}%
  \resizebox{0.23\textwidth}{!}{\includegraphics{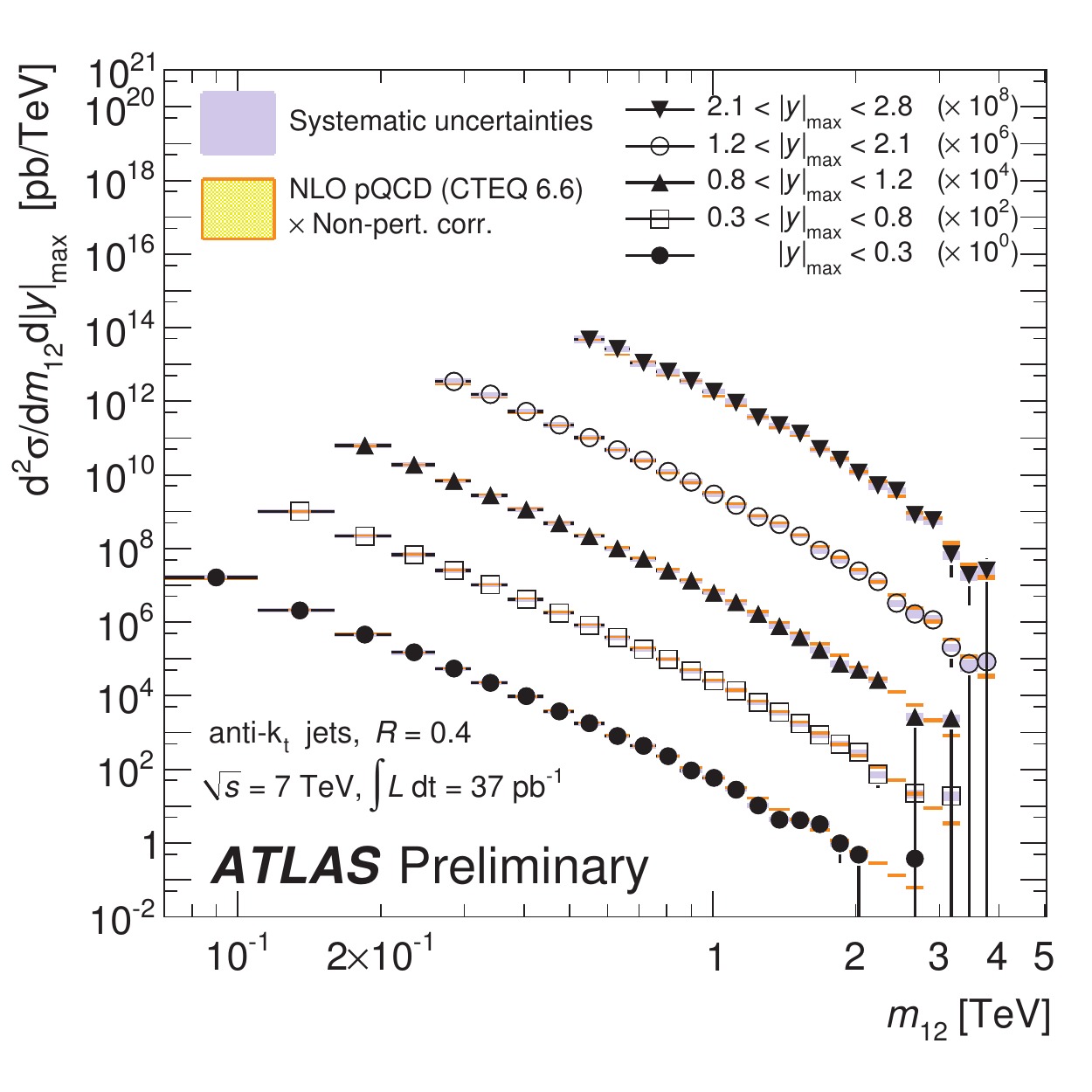}}}
\caption{Inclusive jet cross section from CMS (left) for Anti-$k_\perp$ jets
  with $R=0.5$ as function of $p_\perp$ for $6$ rapidity intervals
  scaled for easier viewing. The data points (symbols) are compared to
  NLO predictions (solid lines) corrected for non-perturbative
  effects. Experimental uncertainties are indicated by the yellow
  bands; Double differential cross section from ATLAS (right) for
  Anti-$k_\perp$ di-jet events with $R=0.4$ as function of di-jet mass
  $m_{12}$ for $5$ intervals of maximum rapidity $|y_{\rm max}|$ with
  systematic experimental uncertainties (grey band). NLO predictions
  with NP corrections and uncertainties are shown as well (yellow
  band).}\label{fig:CMS-incl-xsec:ATLAS-dijet-xsec}
\end{center}
\end{figure}
$p_\perp$ resolution. 
\begin{figure}[htb]
\begin{center}
\centerline{%
  \resizebox{0.23\textwidth}{!}{\includegraphics{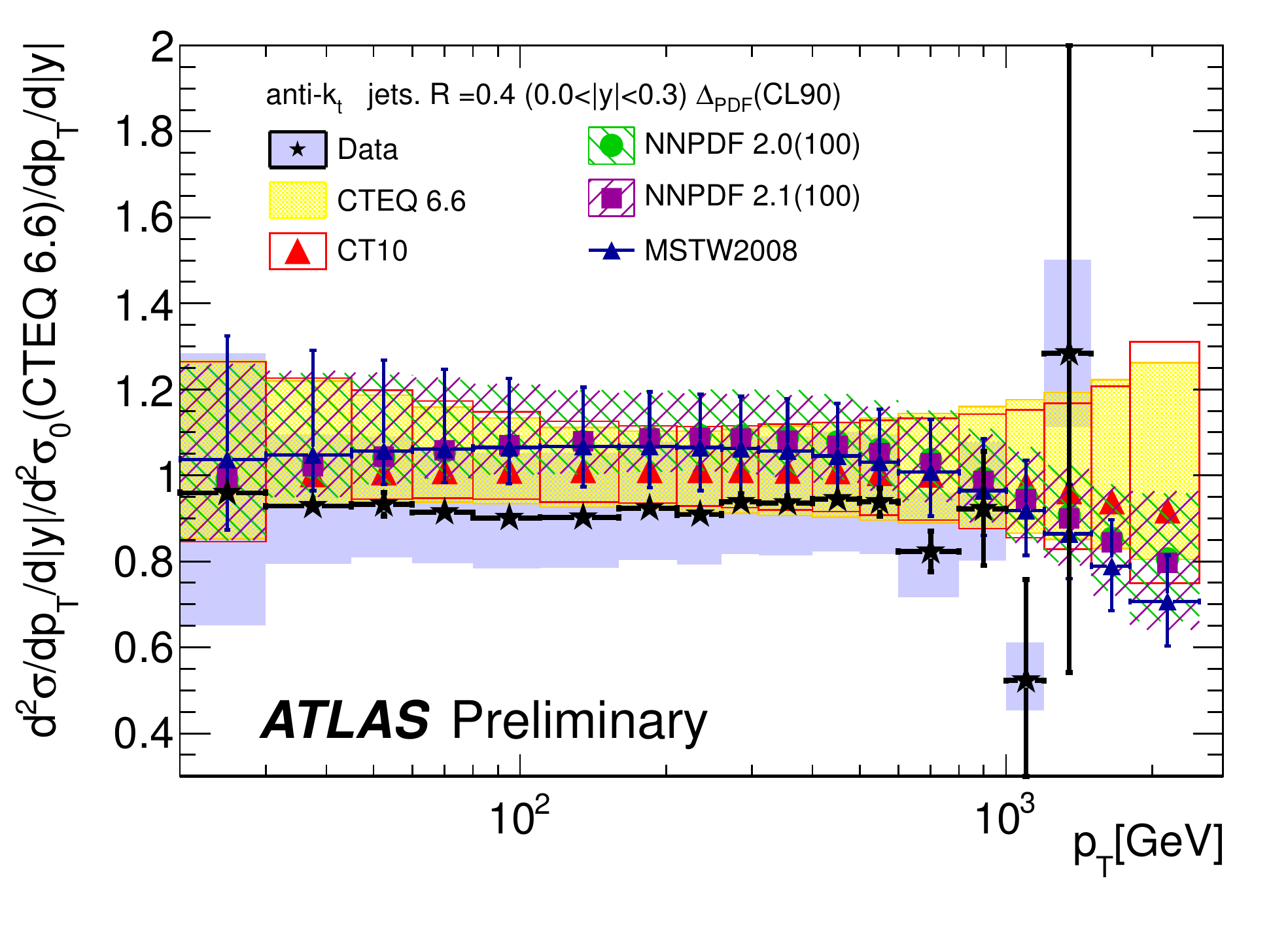}}%
  \resizebox{0.23\textwidth}{!}{\includegraphics{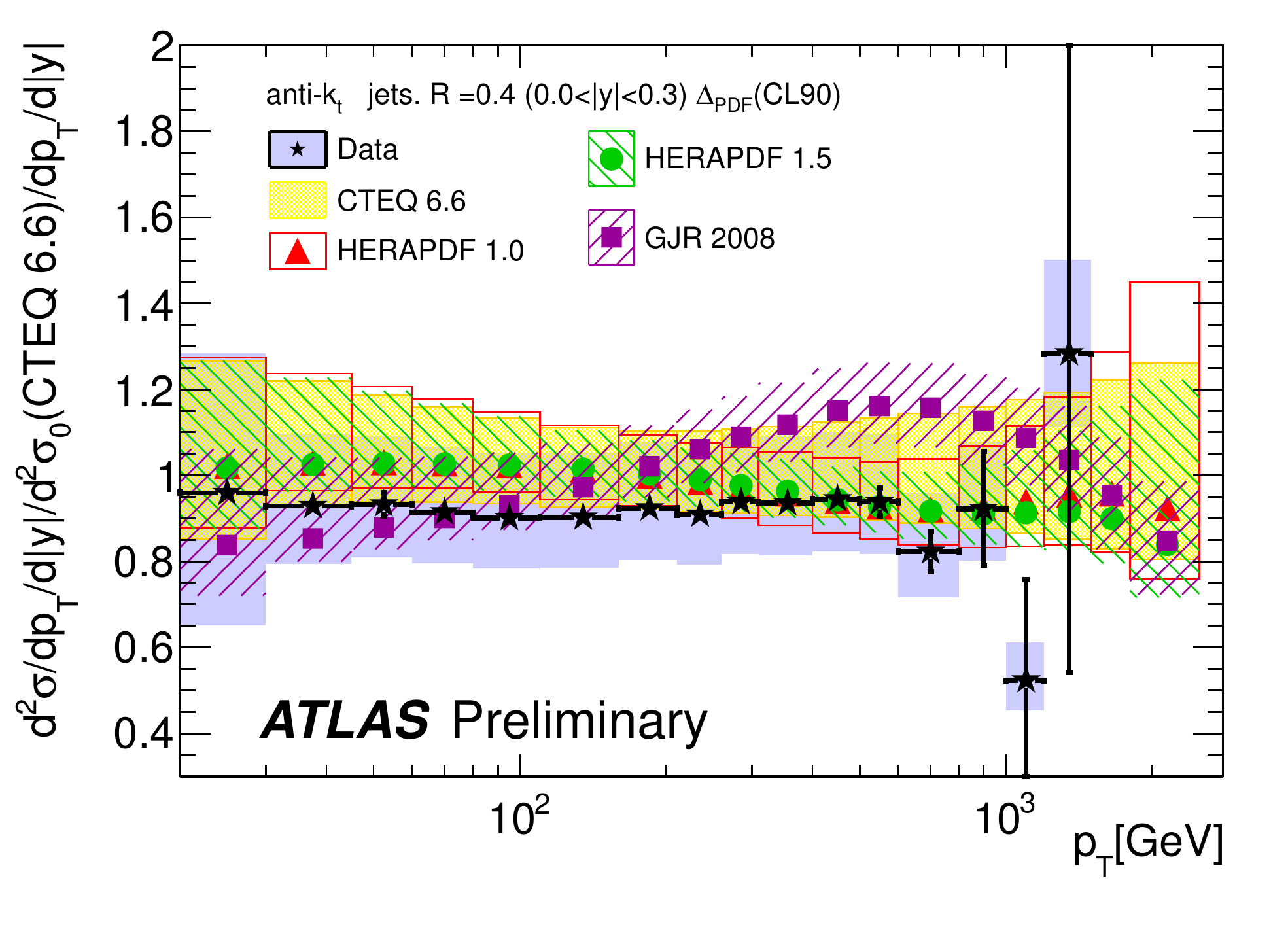}}}
\centerline{%
  \resizebox{0.23\textwidth}{!}{\includegraphics{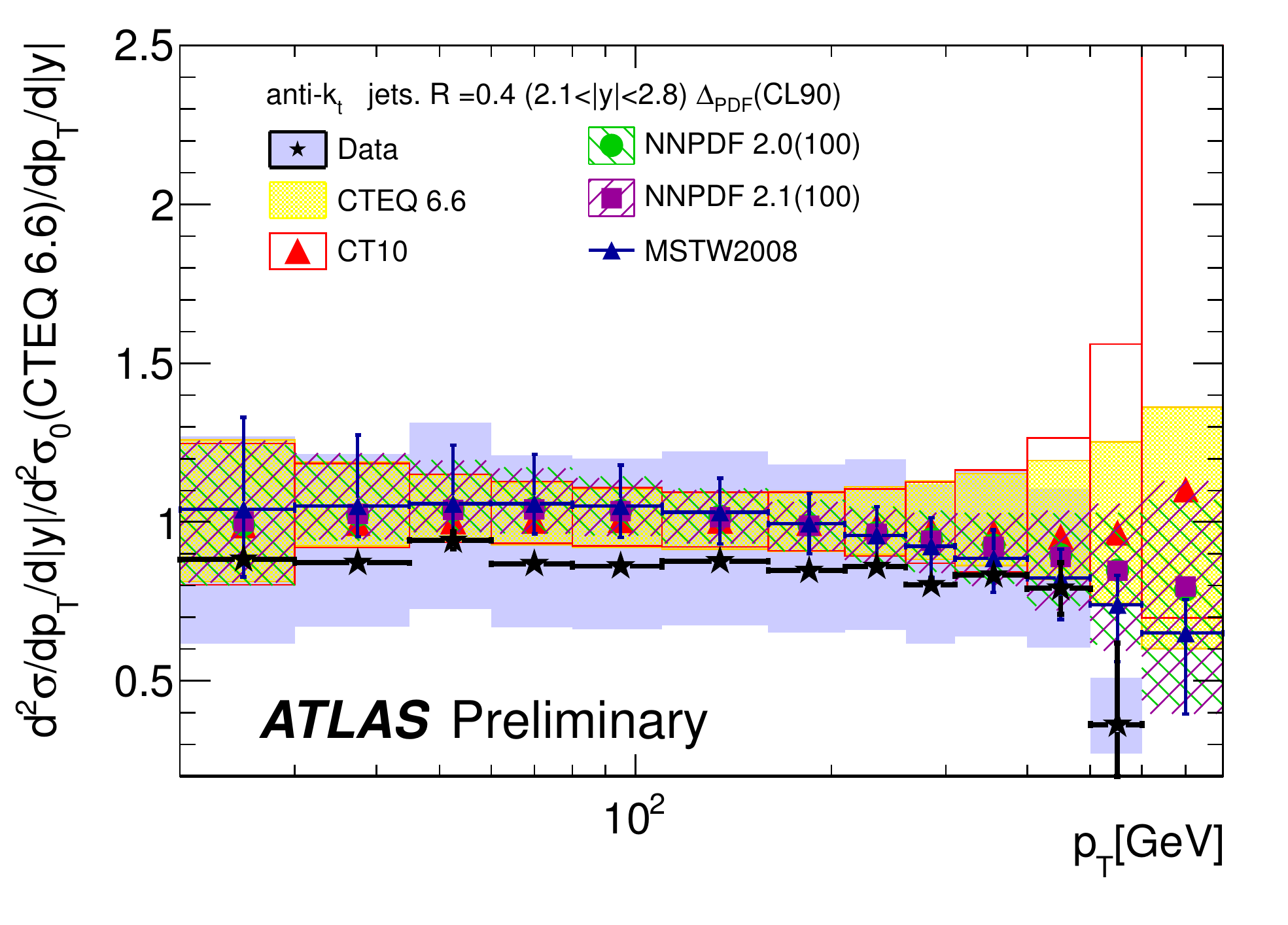}}%
  \resizebox{0.23\textwidth}{!}{\includegraphics{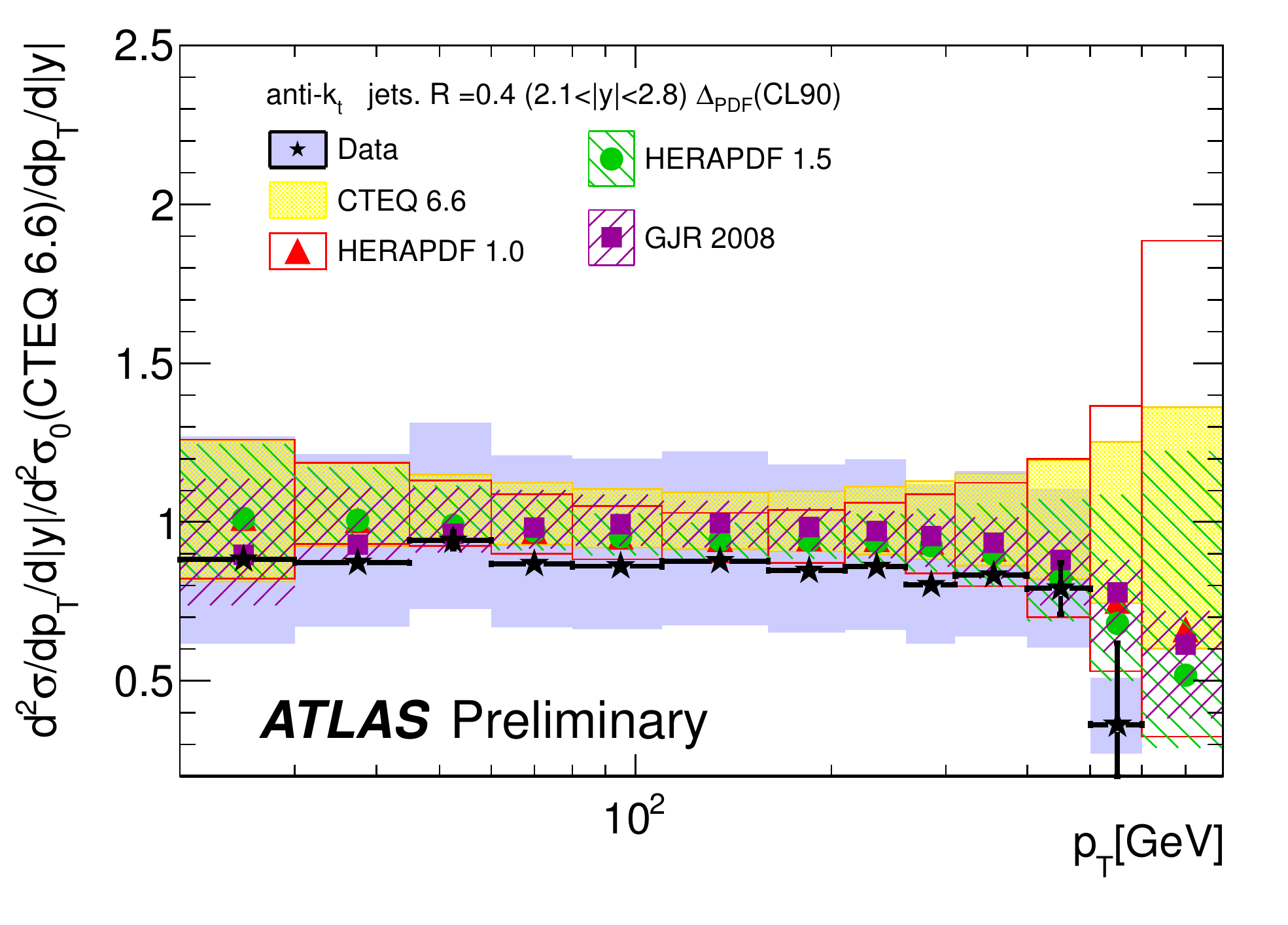}}}
\caption{Ratio of the inclusive jet cross section from ATLAS for
  Anti-$k_\perp$ jets with $R=0.4$ as function of $p_\perp$ over NLO
  predictions %\cite{art:CTEQ,art:CT10,art:MSTW2008,art:NNPDF20,art:NNPDF21a,art:NNPDF21b,art:HERAPDF10,art:HERAPDF15,art:GJR08a,art:GJR08b}
  for $|y|<0.3$ (top) and $2.1 <|y| <2.8$ (bottom).  The reference NLO
  prediction is {\tt CTEQ 6.6} which is compared to the ratios using
  {\tt CT10}, {\tt MSTW 2008}, {\tt NNPDF 2.0} and {\tt NNPDF 2.1}
  (left); {\tt HERAPDF 1.0}, {\tt HERAPDF 1.5} and {\tt GJR08} (right). 
  Error bars indicate statistical errors. The light shaded
  band shows the experimental systematic uncertainties excluding a
  common $3.4\%$ uncertainty from the luminosity measurement. The
  other bands indicate the respective theoretical
  uncertainties.}\label{fig:ATLAS-incl-xsec-ratio}
\end{center}
\end{figure}
In CMS the corrected spectra are obtained by fitting a modified
power-law function with Gaussian smearing in $p_\perp$ to the observed
spectra.  In ATLAS the correction factors are obtained from full
detector simulations including detector inefficiencies. Typical
corrections are in the $10-15\%$ range but can extend to $30-50\%$ at
the edges of the phase space. The NLO perturbative QCD (pQCD)
predictions on parton-level on the other hand are corrected for
non-perturbative (NP) effects due to hadronisation and the underlying
event activities. These corrections are obtained by comparing
simulations with leading log generators ({\tt
  PYTHIA}%~\cite{art:Pythia6}
/{\tt HERWIG}%~\cite{art:HERWIG}
) which are run with and without these effects enabled. The
corrections depend strongly on jet size. For $R=0.5,0.6$ the
underlying event effects dominate and corrections are around $1.2-1.4$
at small $p_\perp$. For $R=0.4$ hadronisation effects are dominant and
corrections of about $0.8$ are obtained at low $p_\perp$. The
corrections approach unity at larger $p_\perp$ for all used $R$
values.

Figure~\ref{fig:CMS-incl-xsec:ATLAS-dijet-xsec} (left) shows the
inclusive jet cross section measurement for jets with size $R=0.5$ as
a function of jet transverse momentum measured by CMS. The
experimental uncertainties are in the range $10-20\%$ and are
dominated by the uncertainties on JES and resolution. Similar
distributions for $R=0.4$ and $R=0.6$ are obtained by ATLAS, with
uncertainties in the range of $10-30\%$.
Different NLO predictions are
tested~\cite{art:ATLAS-PHYS-PUB-2011-005,art:CMS-NOTE-2011-004} by
comparing the ratios of data to NLO MC predictions for various PDF
sets. Figure~\ref{fig:ATLAS-incl-xsec-ratio} shows an example from
ATLAS for the rapidity region $|y|<0.3$ for $R=0.4$ and $R=0.6$. CMS
obtains similar comparisons for $R=0.5$. The NLO predictions are in
general systematically above the data but still compatible with the
measurement within the assigned uncertainties. The deviations become
larger at large $|y|$ and $p_\perp$.
\begin{figure}[htb]
\begin{center}
\centerline{%
  \resizebox{0.23\textwidth}{!}{\includegraphics{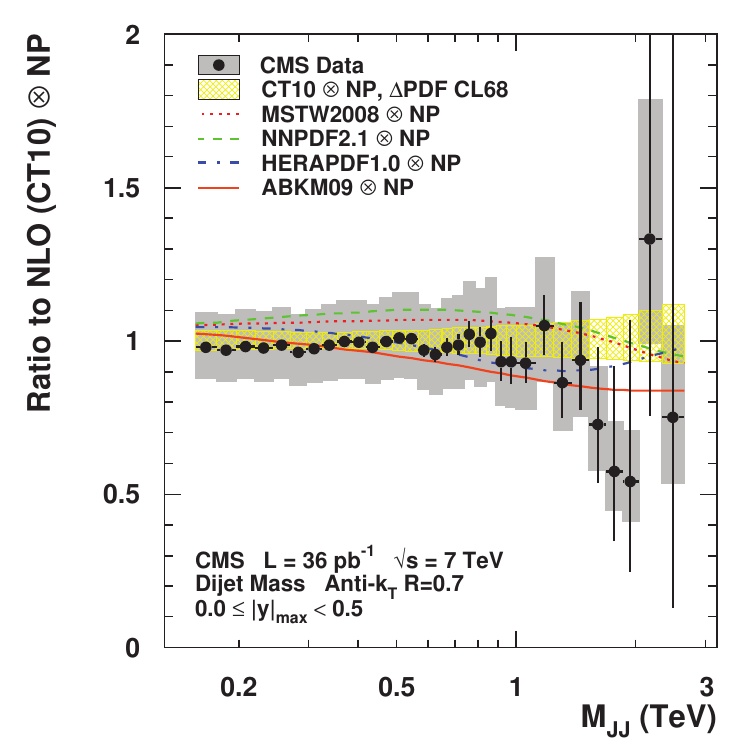}}
  \resizebox{0.23\textwidth}{!}{\includegraphics{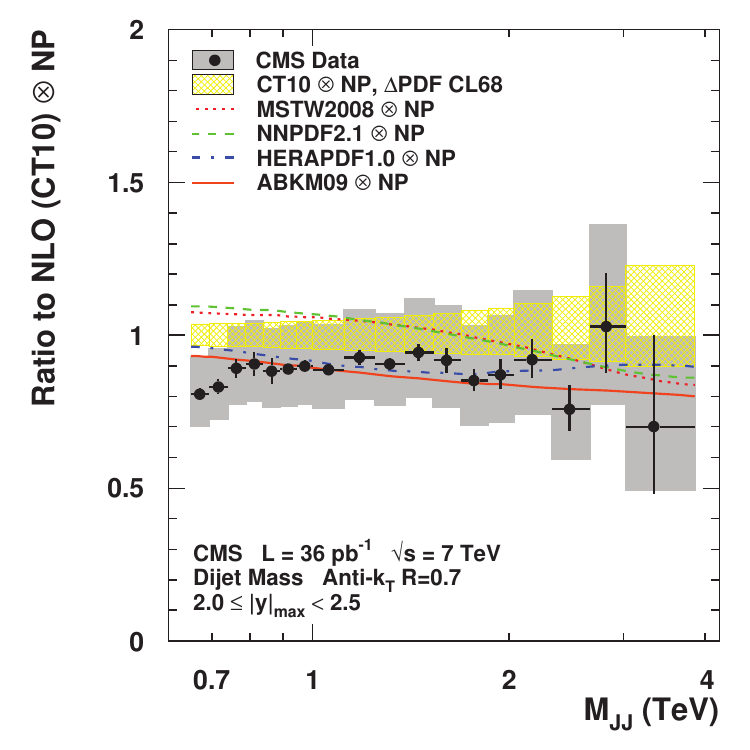}}}
\caption{Ratio of the double differential di-jet cross section from
  CMS for Anti-$k_\perp$ jets with $R=0.7$ as a function of the di-jet
  mass ($M_{\rm jj}$) over NLO
  predictions %\cite{art:CT10,art:MSTW2008,art:NNPDF21a,art:NNPDF21b,art:HERAPDF10,art:ABKM09}
  for $|y|<0.5$ (left) and $2.0 <|y| <2.5$ (right).  The reference NLO
  prediction uses {\tt CT10} which is compared to the ratios using {\tt
    MSTW 2008}, {\tt NNPDF 2.1}, {\tt HERAPDF 1.0} and {\tt ABKM09}
  instead. Error bars indicate statistical errors. The grey band
  shows the experimental systematic uncertainties and the yellow band the typical
  PDF uncertainty ({\tt CT10}). Non-perturbative uncertainties are
  dominant at low masses and not shown in the
  figures.}\label{fig:CMS-dijet-xsec-ratio}
\end{center}
\end{figure}

The double differential cross section in the maximum jet rapidity
$|y_{\rm max}|$ and di-jet mass $m_{12}$ for di-jet events as measured
by ATLAS~\cite{art:ATLAS-CONF-2011-047} is shown in
Figure~\ref{fig:CMS-incl-xsec:ATLAS-dijet-xsec} (right) for
$R=0.4$. Similar results are obtained by CMS~\cite{art:CMS-QCD-10-025}
for $R=0.7$. Both ATLAS and CMS use full simulations to obtain the
bin-by-bin migration corrections for the distributions. Dominant
experimental systematic uncertainties stem from the JES uncertainty
and are in the range of $15-30\%$ for ATLAS and around $15\%$ at low
masses and $60\%$ at high masses for CMS.  As is the case for the
inclusive jet cross section measurement a comprehensive comparison to
NLO pQCD predictions has been made by both
ATLAS~\cite{art:ATLAS-CONF-2011-047} and
CMS~\cite{art:CMS-NOTE-2011-004}. Figure~\ref{fig:CMS-dijet-xsec-ratio}
shows the ratio of the measured double differential di-jet cross
section to that predicted in {\tt CT10}%~\cite{art:CT10}
-based MC simulation for two rapidity bins. The agreement with {\tt
  HERAPDF} is best, but all tested PDF sets agree within
uncertainties.

\section{Angular and multi-jet variables}
\label{Angular}
\begin{figure}[htb]
\begin{center}
\centerline{%
  \resizebox{0.23\textwidth}{!}{\includegraphics{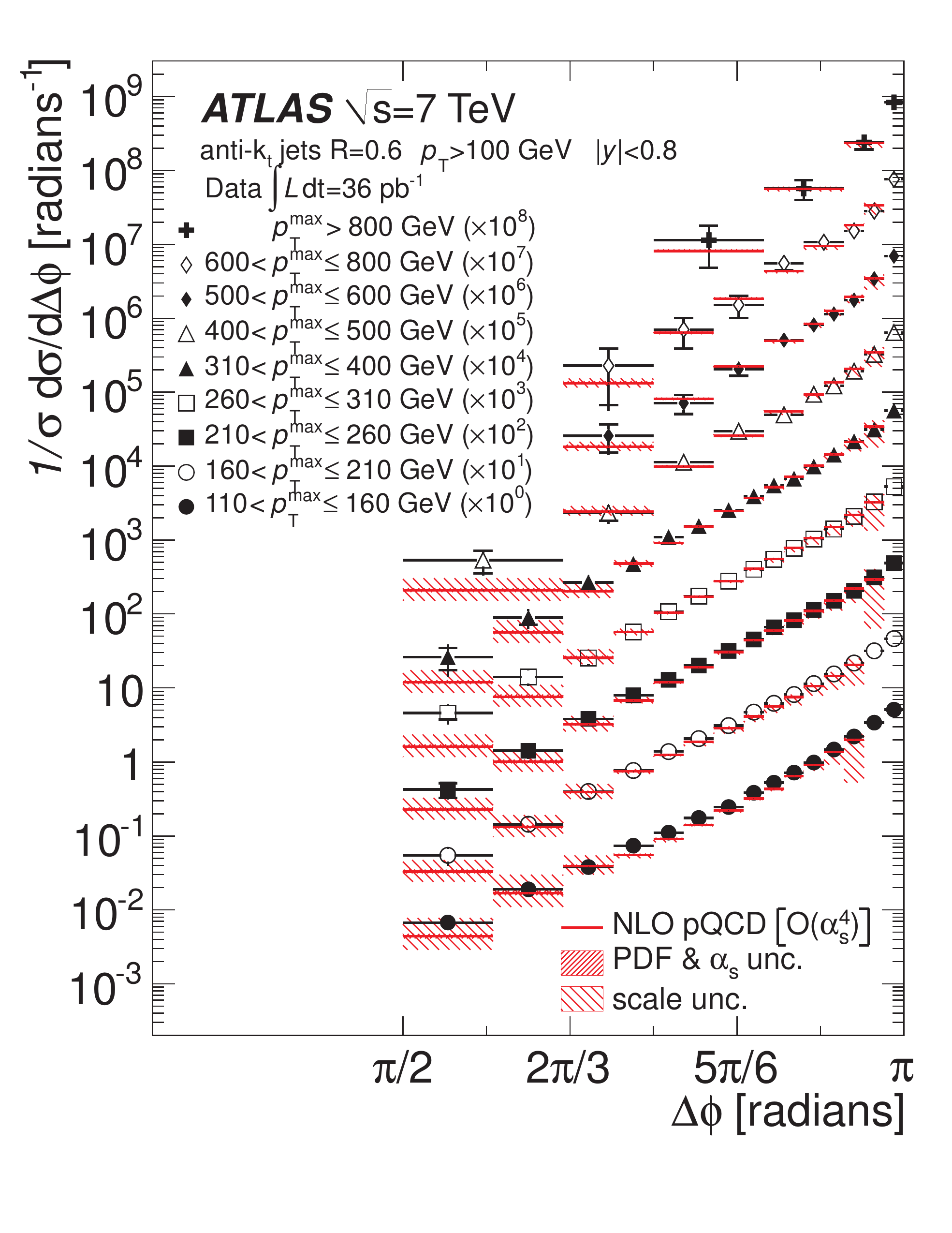}}%
  \resizebox{0.24\textwidth}{!}{\includegraphics{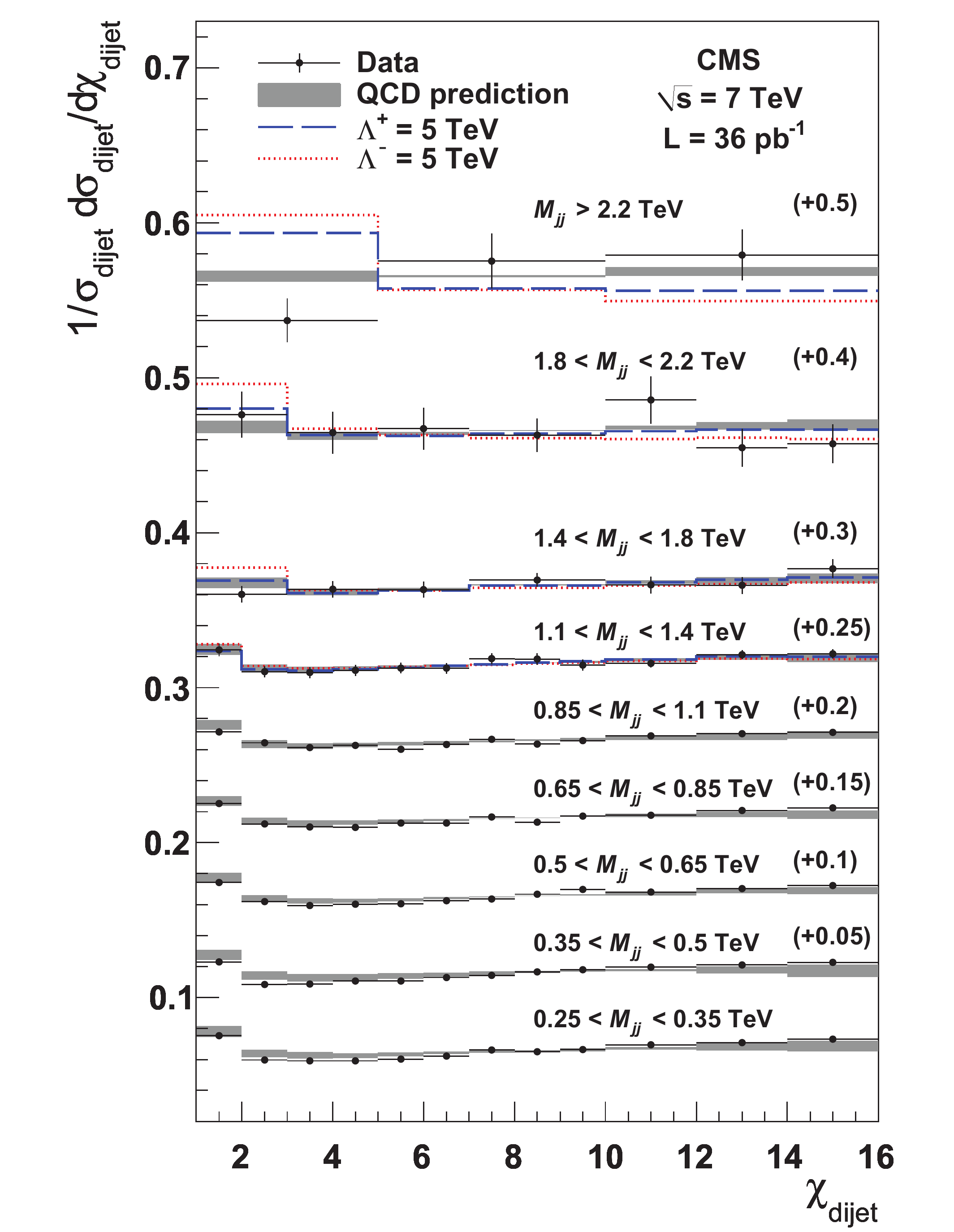}}}
\caption{Azimuthal de-correlation $\Delta\phi$ of the two most
  energetic jets as measured by ATLAS (left) for Anti-$k_\perp$ jets
  with $R=0.6$ for different $p_\perp^{\rm max}$ intervals (black
  markers) compared to NLO pQCD calculations (red lines) with
  associated errors (hatch pattern); The distribution of $\chi_{\rm
    dijet}$ (black points) for different $M_{\rm jj}$ ranges as
  measured by CMS (right) for Anti-$k_\perp$ jets with $R=0.5$
  compared to NLO pQCD calculations (shaded band) and predictions
  including contact interactions (colored lines) for compositeness
  scales of $\Lambda^{+/-} = 5\,{\rm
    TeV}$.}\label{fig:ATLAS-Delta-phi:CMS-chi}
\end{center}
\end{figure}

\begin{figure}[htb]
\begin{center}
\centerline{%
  \resizebox{0.2\textwidth}{!}{\includegraphics{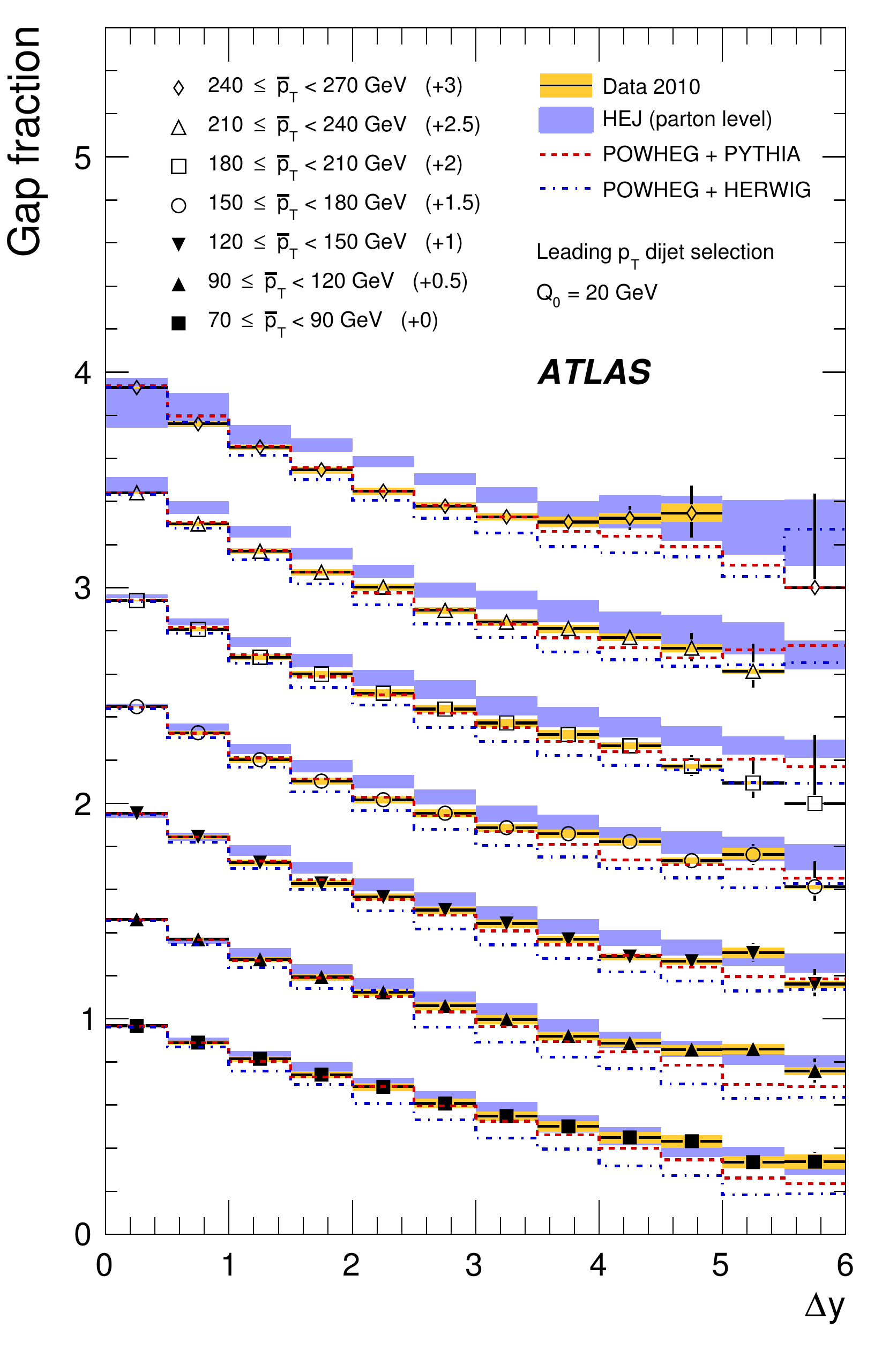}}%
  \resizebox{0.29\textwidth}{!}{\includegraphics{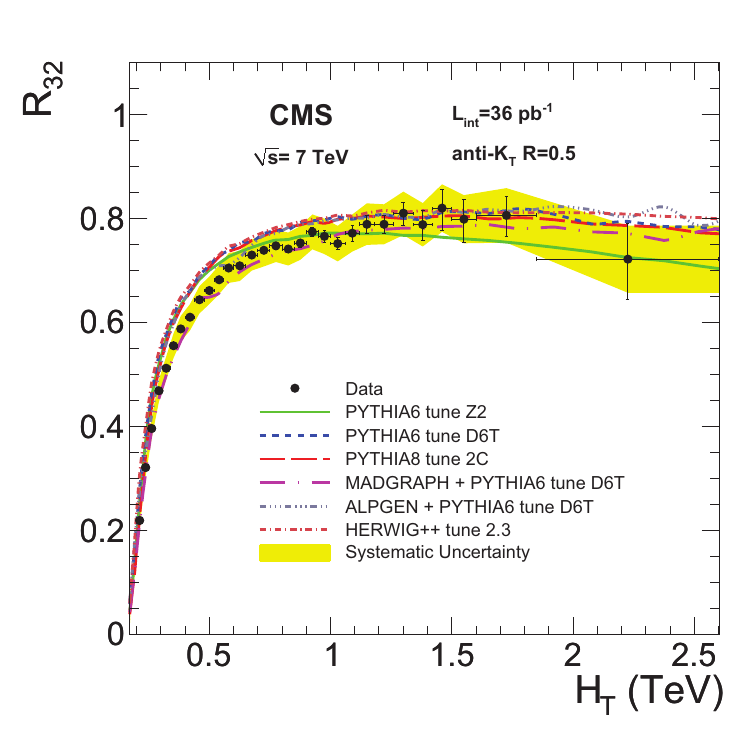}}}
\caption{Gap-fraction (left) for the two leading jets in $p_\perp$ as
  a function of $\Delta y$ for various intervals of $\bar{p}_\perp$ as
  measured by ATLAS (markers and yellow error band) compared to {\tt
    HEJ} calculations (blue band) and NLO simulations with {\tt
    POWHEG} interfaced to {\tt PYTHIA} (red dashes) and {\tt HERWIG}
  (blue dot-dashes); Ratio of inclusive $3$-jet over the $2$-jet cross
  sections $R_{32}$ as function of total transverse momentum $H_\perp$
  (right) as measured by CMS (black dots and yellow uncertainty band)
  compared to various simulations using different {\tt PYTHIA} tunes,
  {\tt MADGRAPH}, {\tt ALPGEN} and {\tt
    HERWIG++}.}\label{fig:ATLAS-gap-fraction:CMS-R32}
\end{center}
\end{figure}

Due to their sensitivity to new physics and their ability to probe
mass scales without explicitly relying on JES calibrations the angular
distributions of multi-jet events are of particular interest. The
azimuthal de-correlation $\Delta\phi$ of the two most energetic jets
as measured by ATLAS~\cite{art:ATLAS-DiJet-DeltaPhi} is shown in the
left plot of figure~\ref{fig:ATLAS-Delta-phi:CMS-chi}. Values close to
$\pi$ are expected for di-jet events while smaller values indicate the
presence of additional jets. NLO pQCD calculations using {\tt
  NLOJet++} %\cite{art:NLOJet} 
and {\tt MSTW 2008} %\cite{art:MSTW2008}
agree with the data for $\Delta\phi < \pi$. Leading log simulations
({\tt PYTHIA}, {\tt HERWIG}, {\tt SHERPA}
%~\cite{art:SHERPA})
agree with the data and give a good description of the perturbatively diverging
point $\Delta\phi = \pi$. The right side plot of
figure~\ref{fig:ATLAS-Delta-phi:CMS-chi} shows the distribution of
$\chi_{\rm dijet} = \exp{|y_1 - y_2|}$, the exponential of the
rapidity difference between the two leading jets in $p_\perp$, as
measured by CMS~\cite{art:CMS-dijet-angular} for different di-jet mass
intervals. The distribution in $\chi_{\rm dijet}$ is expected to be
almost flat for QCD while new physics (such as quark compositeness) would
cause excess events at small $\chi_{\rm dijet}$. The comparison to NLO
pQCD calculations with {\tt NLOJet++} %\cite{art:NLOJet}
and the {\tt CTEQ 6.6} %\cite{art:CTEQ}
PDF-set shows good agreement with the data, and a lower limit on the
contact interaction scale for left-handed quarks of
$\Lambda^+=5.6\,{\rm TeV}$ ($\Lambda^-=6.7\,{\rm TeV}$) for
destructive (constructive) interference has been obtained at $95\%$
CL. A complementary study of the rapidity gap between the two jets with
either leading $p_\perp$ or the largest rapidity gap $\Delta y$ has
been done by ATLAS~\cite{art:ATLAS-rapidity-gap}. The so-called
gap-fraction is defined as the fraction of events without additional
jet activity in the rapidity interval between the two jets. Any
additional jet within the gap has to have a transverse momentum above
a veto scale $p_\perp > Q_0$, with the default choice $Q_0 =
20\,{\rm GeV}$ to stay far away from $\Lambda_{\rm QCD}$. The
gap-fraction is shown in the left plot of
figure~\ref{fig:ATLAS-gap-fraction:CMS-R32} for the choice of leading
jets in $p_\perp$ as a function of $\Delta y$ for various intervals of
the average transverse momentum of the two leading jets
$\bar{p}_\perp$. The comparison with HEJ %\cite{art:HEJ}
calculations shows some deviations in the large $\bar{p}_\perp$
regions but the agreement improves as $\bar{p}_\perp$ approaches 
$Q_0$, which is expected since HEJ is designed to give a
good description of QCD in the limit where all jets have similar
$p_\perp$. The best description is achieved with {\tt
  POWHEG} %\cite{art:POWHEGa,art:POWHEGb}
interfaced to {\tt PYTHIA} %\cite{art:Pythia6}
although deviations are observed at large $\Delta y$. {\tt POWHEG}
interfaced to {\tt HERWIG} %\cite{art:HERWIG}
tends to predict smaller gap fractions over the full phase space and
the deviations increase for larger $\Delta y$ as for the {\tt
  POWHEG}+{\tt PYTHIA} case.

The right hand plot in figure~\ref{fig:ATLAS-gap-fraction:CMS-R32}
shows the cross section ratio of three-jet over two-jet events
$R_{32}$ as a function of the total transverse momentum sum $H_\perp =
\sum_{\rm jets}{p_\perp}$ as measured by CMS~\cite{art:CMS-R32}. Many
systematic uncertainties such as those due to the JES and the jet
selection efficiency largely cancel in this ratio, while the
uncertainty due to the integrated luminosity vanishes
entirely. Therefore $R_{32}$ provides a stringent test of QCD
predictions. Events with two or more Anti-$k_\perp$ jets with $R=0.5$
with $|y|<2.5$ and $p_\perp > 50\,{\rm GeV}$ and $H_\perp > 0.2\,{\rm
  TeV}$ are selected and compared to various {\tt PYTHIA6}, {\tt
  PYTHIA8} and {\tt HERWIG++} based tunes and to simulations using the
multi-parton final state generators {\tt MADPGRAPH} and {\tt ALPGEN}
interfaced to {\tt PYTHIA6}. All predictions describe the observed
ratio well in the region $H_\perp > 0.5\,{\rm TeV}$ but, with the
exception of {\tt MADPGRAPH}, overshoot between $10-30\%$ at lower
$H_\perp$. 
 
\section{Jet mass and sub-structure}
\label{JetSub}
In the high energy regime of LHC, heavy objects with masses $O(100\,{\rm GeV})$, can receive large Lorentz boosts
such that their decay products are measured in a single jet. Several approaches
are considered to explore the sub-structure of these jets with the aim
to identify such heavy objects. Among them are:
\begin{description}
\item[{\bf C/A filtering}:] The clustering of large ($R\simeq1.2$)
  Cambridge-Aachen (C/A)~\cite{art:CambridgeAachen} type jets is
  reversed until a large drop in jet-mass is observed. The remaining
  constituents are re-clustered with a smaller $R$ parameter.
\item[{\bf Jet pruning}:] C/A or $k_\perp$~\cite{art:Kta,art:Ktb}
  jet-clustering is performed on the constituents of a large jet and
  in each clustering step the softer of the two clusters being
  combined is discarded if it's transverse momentum is below a certain
  fraction of the original jet $p_\perp$ and the angular distance
  between the two clusters is large.
\end{description}
For the jet sub structure algorithms to be useful they have to be
tested on QCD jets as this will be the main background.  C/A Filtering
is useful for the decays of heavy particles to two low mass objects
and the QCD behavior has been studied in ATLAS
in~\cite{art:ATLAS-CONF-2011-073}. The mass drop $m_1/m_{\rm jet}$ of
the leading subjet is required to be smaller than $0.67$ (light
subjet) and the $p_\perp$ asymmetry $\left({\rm
    min}(p_\perp^1,p_\perp^2)\times\Delta R_{1,2}/m_{\rm
    jet}\right)^2$ larger than $0.09$ (fairly symmetric). Once a
reversed clustering step with these properties is found the current
jet is re-clustered with C/A and $R = {\rm min}(0.3, \Delta
R_{1,2}/2)$ finding $n$ new subjets of which the leading ${\rm
  min}(3,n)$ are combined to give the final C/A filtered
jet. 
\begin{figure}[htb]
\begin{center}
\centerline{%
  \resizebox{0.23\textwidth}{!}{\includegraphics{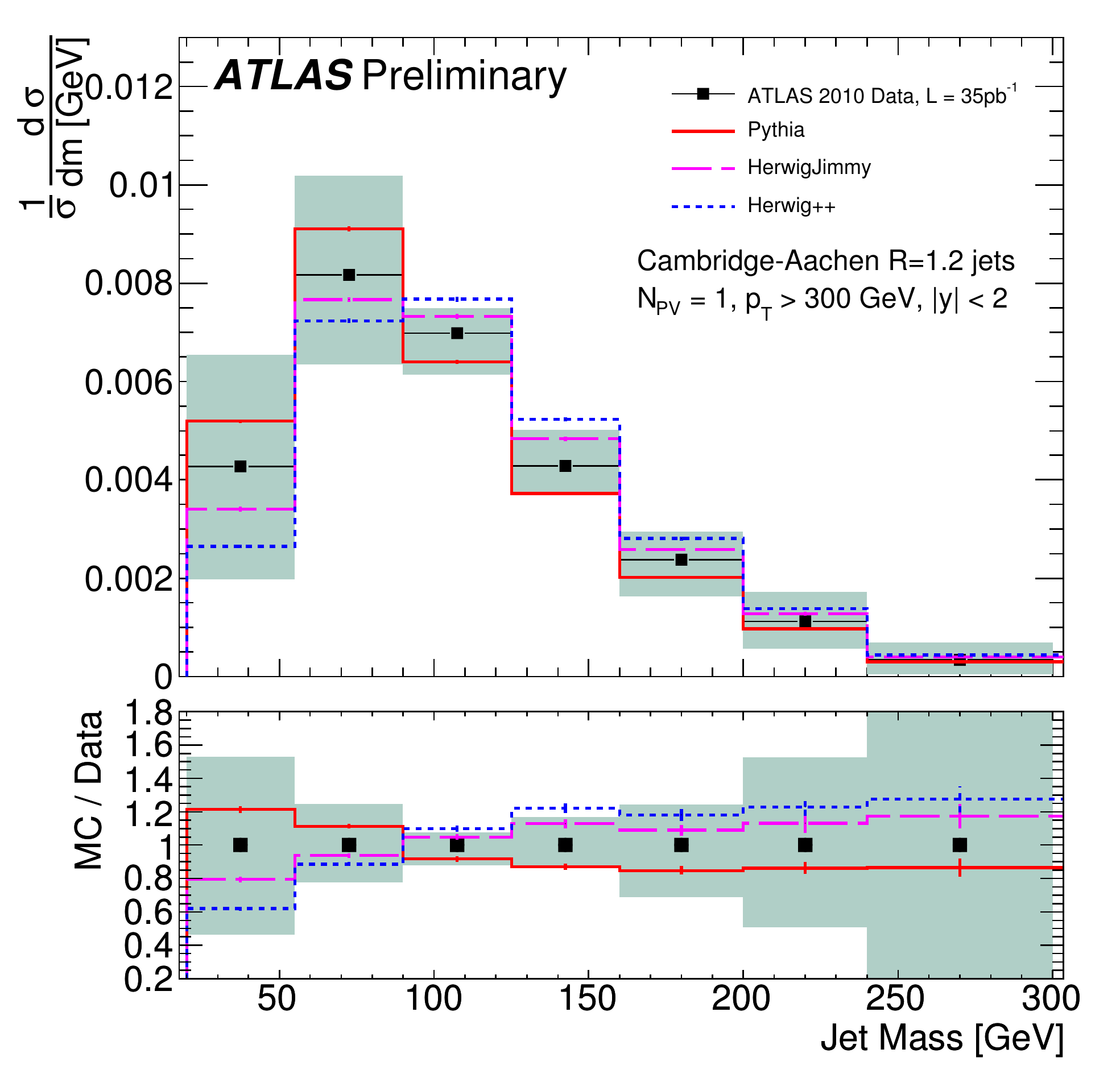}}%
  \resizebox{0.23\textwidth}{!}{\includegraphics{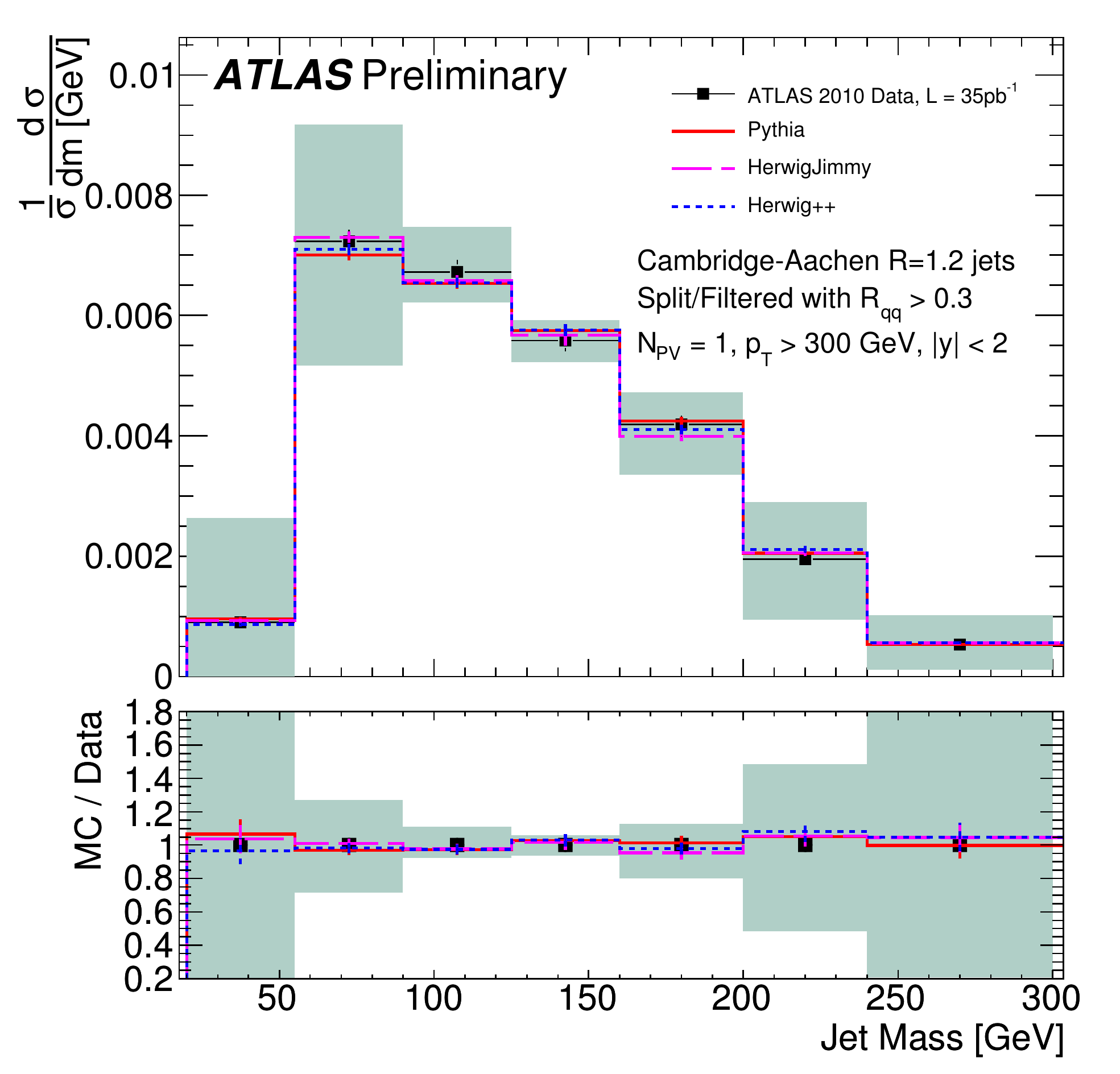}}}
\caption{Jet masses for C/A jets with $R=1.2$ before (left) and after
  (right) the filtering procedure (see text) as measured by ATLAS. The
  data (black points and shaded error band) is fully corrected for
  detector effects and compared to {\tt PYTHIA}, {\tt HERWIG/JIMMY}
  and {\tt HERWIG++}. The lower portions of the plots show ratios of
  the distributions over data.}\label{fig:ATLAS-Jet-Mass}
\end{center}
\end{figure}
Figure~\ref{fig:ATLAS-Jet-Mass} shows the spectrum of jet masses
for C/A jets with $R=1.2$ before and after the filtering procedure in
events with exactly one primary vertex (to remove pile-up) and at
least one jet with $p_\perp > 300\,{\rm GeV}$ and $|y|<2$.  The
agreement with all three predictions is good although {\tt HERWIG++}
produces jets with larger mass (before filtering) compared to data.
%After filtering the simulations are in much better agreement
%with each other and with data.
\begin{figure}[htb]
\begin{center}
\centerline{%
  \resizebox{0.23\textwidth}{!}{\rotatebox{90}{%
      \includegraphics{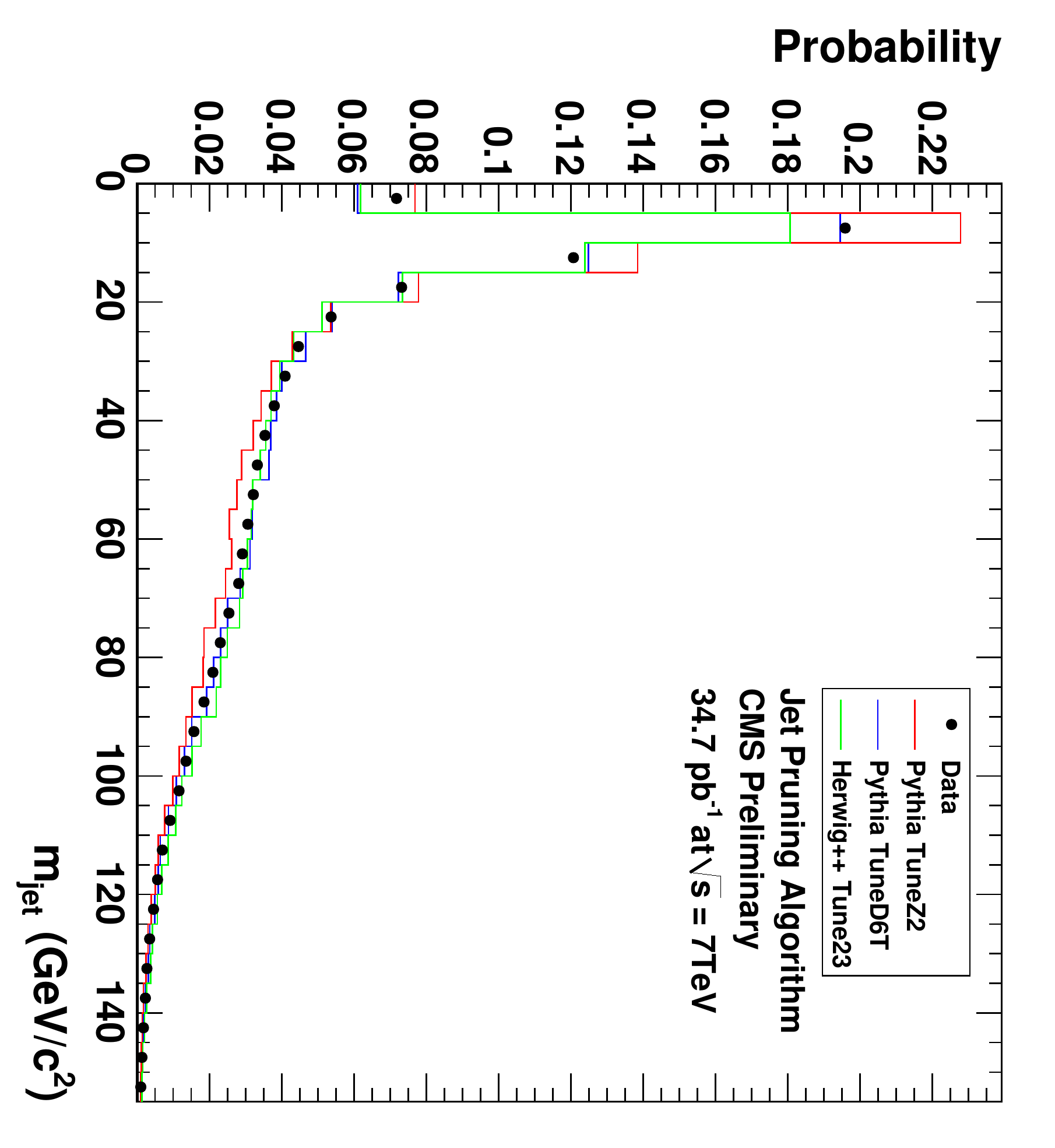}}}%
  \resizebox{0.23\textwidth}{!}{\rotatebox{90}{%
      \includegraphics{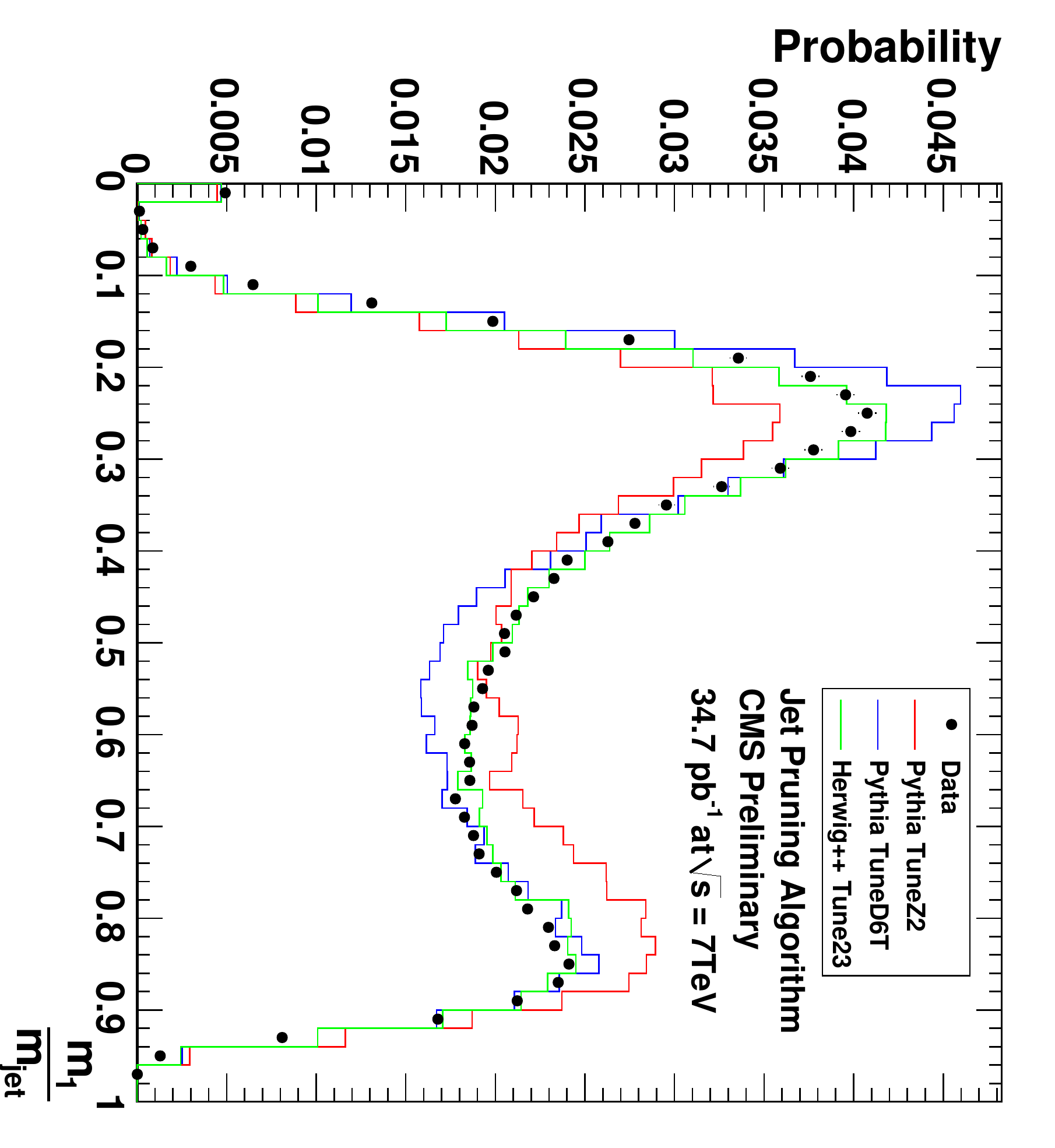}}}}
%\centerline{%
%  \resizebox{0.23\textwidth}{!}{\rotatebox{90}{%
%      \includegraphics{../CMS/JetShape_Figure_009-c}}}%
%  \resizebox{0.23\textwidth}{!}{\rotatebox{90}{%
%      \includegraphics{../CMS/JetShape_Figure_009-d}}}}
\caption{Leading jet pruning quantities in events with at least $2$
  hard jets as measured by CMS (data points). Shown are the mass of
  the pruned jet (left) and the mass drop of the leading subjet (right)
  % ,subjet $\Delta R$ (bottom left) and $p_\perp$ asymmetry
  % $y$ (bottom right).
  The data is normalized to unity
  and compared to two {\tt PYTHIA} tunes (red and blue) and {\tt
    HERWIG++} (green).}\label{fig:CMS-pruned-jets}
\end{center}
\end{figure}
Jet pruning is useful for $\rm W$ tagging and the QCD behavior has
been tested by CMS in~\cite{art:CMS-PAS-JME-10-013}.  For boosted $\rm
W$s decaying into two quarks with similar energy and mass two light
subjets are expected in the pruning algorithm with the pruned jet mass
close to $m_{\rm W}$. The mass drop $m_1/m_{\rm jet}$ of the leading
subjet should be smaller than $0.4$ consistent with two light subjets.
Figure~\ref{fig:CMS-pruned-jets} shows the pruning properties of the
leading jet in events with at least two high $p_\perp > 200\,{\rm
  GeV}$ jets with $\Delta\phi > 2.1$ and $|\eta| < 2.5$ in comparison
to two different {\tt PYTHIA} tunes and {\tt HERWIG++}. The overall
agreement of the data with simulation is good -- especially with the
{\tt HERWIG++} tune.
\section{Conclusions}
\label{Conclusions}
Both ATLAS and CMS have made comprehensive studies of hard QCD
involving jets. Excellent agreement with NLO pQCD calculations has
been found and constraints on new physics were set by the observed
agreement. Novel techniques to identify massive boosted objects were
successfully tested on the large QCD background expected. The
challenge will be to continue the studies presented here under the
increased pile-up conditions in the data taken beyond 2010.
\section*{Acknowledgments}
\label{Acknowledgments}
I'd like to thank the Jet Performance and SM/QCD groups of ATLAS and
CMS for providing me with the material presented here. In particular I
benefited greatly from discussions with J.~Butterworth, M.~Campanelli,
A.~Davison, A.~Di Ciaccio, K.~Kousouris, and M.~Voutilainen.
\bibliographystyle{epj}
\bibliography{hcp2011proc}
\end{document}